\input{aipcheck}

\documentclass[
  ,draft            
  ]
  {aipproc}

\layoutstyle{8x11single}

\def\IZ{\hbox{\sf Z\kern-.4em Z}}
\def\IR{\hbox{\rm I\kern-.18em R}}
\def\IC{\hbox{\,$\inbar\kern-.3em{\rm C}$}}
\def\one{\hbox{{1}\kern-.25em\hbox{l}}}
\begin{document}

\title{The concept of quasi-integrability}

\classification{05.45.Yv, 81.07.De, 63.20.K-,11.10.Lm, 11.27.+d}

\keywords{integrability, integrable field theories, solitons}

\author{Luiz. A. Ferreira}{
  address={Instituto de F\'\i sica de S\~ao Carlos; IFSC/USP;\\
Universidade de S\~ao Paulo  \\ 
Caixa Postal 369, CEP 13560-970, S\~ao Carlos-SP, Brazil\\ laf@ifsc.usp.br}
}

\author{G. Luchini}{
address={Departamento de Ci\^encias Naturais; CEUNES;
Universidade Federal do Esp\'irito Santo \\
CEP 29932-540, S\~ao Mateus-ES, Brazil \\ gabriel.luchini@ceunes.ufes.br}
}
\author{Wojtek J. Zakrzewski}{
  address={Department of Mathematical Sciences,\\
 University of Durham, Durham DH1 3LE, U.K.\\ w.j.zakrzewski@durham.ac.uk}
}

\begin{abstract}
We show that certain field theory models, although non-integrable according to the usual definition of integrability, share some of the features of integrable theories for certain configurations. Here we discuss our attempt to define a ``\emph{quasi}-integrable theory'', through a concrete example: a deformation of the (integrable) sine-Gordon potential. The techniques used to describe and define this concept are both analytical and numerical. The zero-curvature representation and the abelianisation procedure commonly used in integrable field theories are adapted to this new case and we show that they produce asymptotically conserved charges that can then be observed in the simulations of scattering of solitons.
\end{abstract}

\maketitle


\section{Introduction}

The integrability of a finite degree of freedom dynamical system is related to the fact that the system possesses as many conserved quantities as its number of degrees of freedom. For field theories this concept is much harder to define since the number of degrees of freedom is infinite: having an infinite number of conserved charges does not imply that we have them all. Although the real physical World is certainly non-integrable, integrable models are important as they can be used as approximations to the description of many real phenomena with good accuracy. 

The standard definition of integrability for field theories in $(1+1)$ dimensions is usually provided in terms of the existence of the Lax pair or the so-called Lax-Zakharov-Shabat equation \cite{LAX}\cite{LZS}, which is then shown to produce the infinite number of conserved quantities responsible for the stability of soliton solutions in these theories. However, it has also been observed that some non-integrable models can also possess soliton-like configurations which behave not very differently from solitons in integrable models. This
has been observed even in (2+1) dimensions; the examples here are {\it e.g.} the Ward modified chiral models and the baby Skyrme models with many potentials. 

These facts suggest the following question: can we go beyond integrability and define quasi-integrability (or almost integrability)?
And if we can would it be relevant for all configurations or just for some very special ones? Is there anything else we could say about such ``almost'' integrable models? Is it possible to quantify how integrable such  system are?

This is certainly not the first time such questions have been formulated and is certainly not the last one. We do not have an answer for our questions but we do have some clues and these clues are based on some concrete examples. Here we will discuss one of them; based on what was observed we have formulated our first definition of a quasi-integrable model.\cite{SG} and \cite{NLS}  Our definition can be summarized as follows. 

Consider a $(1+1)$ dimensional field theory with soliton-like solutions that has an infinite number of quasi-conserved quantities $Q_n$, $n\in \IZ$,
$$
\frac{dQ_n}{dt}=\alpha_n.
$$  
The $\alpha_n$ is called an {\bf anomaly} and it vanishes when the theory is integrable for all field configurations. Moreover, it also 
vanishes in  an quasi-integrable theory but only for one-soliton static configutations. However, when one considers $\alpha_n$ for
two solitons configurations in these theories the anomaly has the intriguing property 
$$
\int_{-\infty}^{+\infty}dt\;\alpha_n = 0,
$$
which implies that the charges are asymptotically conserved:
$$
Q_n(t=-\infty)=Q_n(t=+\infty).
$$
If the model possesses breather-like field configurations we have instead
$$
\int_0^Tdt\;\alpha_n=0 \qquad \Rightarrow \qquad Q_n(t)=Q_n(t+T).
$$
where $T$ is some period defined by the particular breather solution under consideration. 

\section{A concrete example}

The first model where the properties above mentioned  were first observed was the scalar field theory with potential
\begin{equation}
\label{pot_bazeia}
V(\varphi,\varepsilon)=\frac{2}{\left(2+\varepsilon\right)^2}\tan^2(\varphi)\left(1-\vert \sin \varphi\vert^{2+\varepsilon}\right)^2,
\end{equation}
with $\varepsilon$ a parameter defining a deformation from the sine-Gordon potential, that can be recovered by taking $\varepsilon=0$: $V(\varphi,0)=\frac{1}{16}\left(1-\cos(4\varphi)\right)$.

\begin{figure}[!ht]
\centering
\includegraphics[scale=0.7]{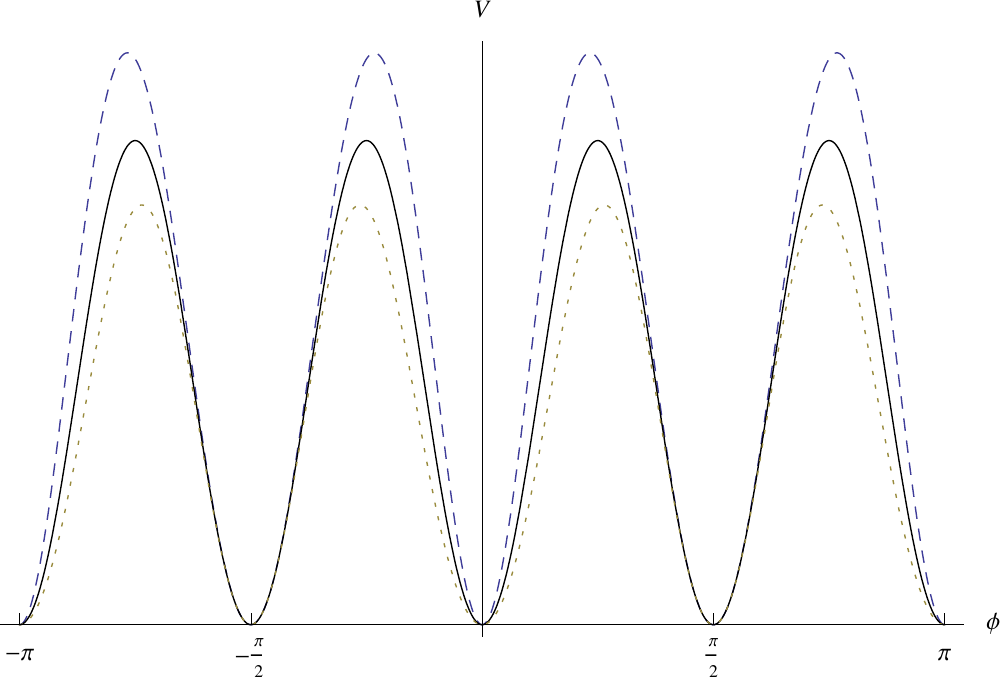}
\caption{The potential for different values of $\varepsilon$. The dashed line is for $\varepsilon=-0.5$, the full line is for $\varepsilon=0.0$ and the dotted line for $\varepsilon=0.5$. }
\label{fig:potential}
\end{figure}

The infinite number of degenerate vacua of the theory allows for the existence of kink solutions (satisfying Bogomolnyi bound) given by
\begin{equation}
\label{kink_bazeia}
\varphi = \pm \arcsin\left(\frac{e^{2\Gamma}}{1+e^{2\Gamma}}\right)^{\frac{1}{2+\varepsilon}} + l\, \pi,\qquad\qquad\qquad \Gamma = \pm \frac{(x-x_0-vt)}{\sqrt{1-v^2}}.
\end{equation}

\begin{figure}[!ht]
\centering
\includegraphics[scale=0.7]{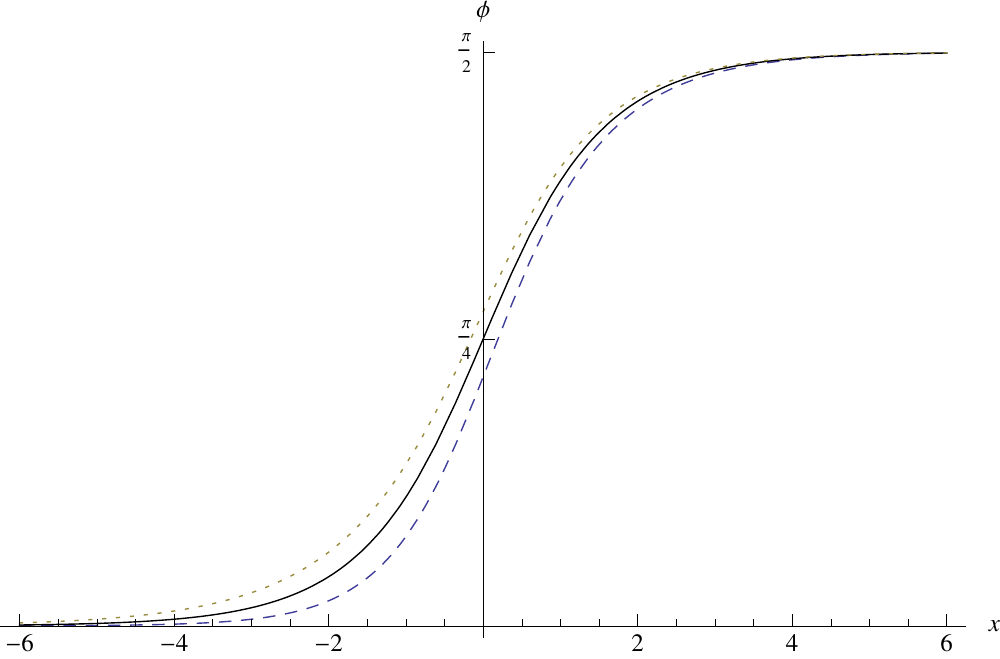}
\caption{The vacua are independent of $\varepsilon$ and therefore they are the same as in the sine-Gordon theory.}
\label{fig:kink}
\end{figure}

Note that when $\varepsilon\neq 0$ there are two types of kinks, of topological charge $+1$ and two types of anti-kinks, of topological charge $-1$. The topological charges are given by the product of the factors $\pm 1$ in $\Gamma$ and in $\varphi$ above. The existence of two classes of kinks and anti-kinks is due to the fact that for $\varepsilon\neq 0$ the discrete symmetry of the potential is $\varphi \rightarrow \varphi + l\pi$, $l\in \IZ$ and not $\varphi \rightarrow \varphi + l\frac{\pi}{2}$, as in the $\varepsilon=0$ case.

\begin{figure}[!ht]
\centering
\includegraphics[scale=0.7]{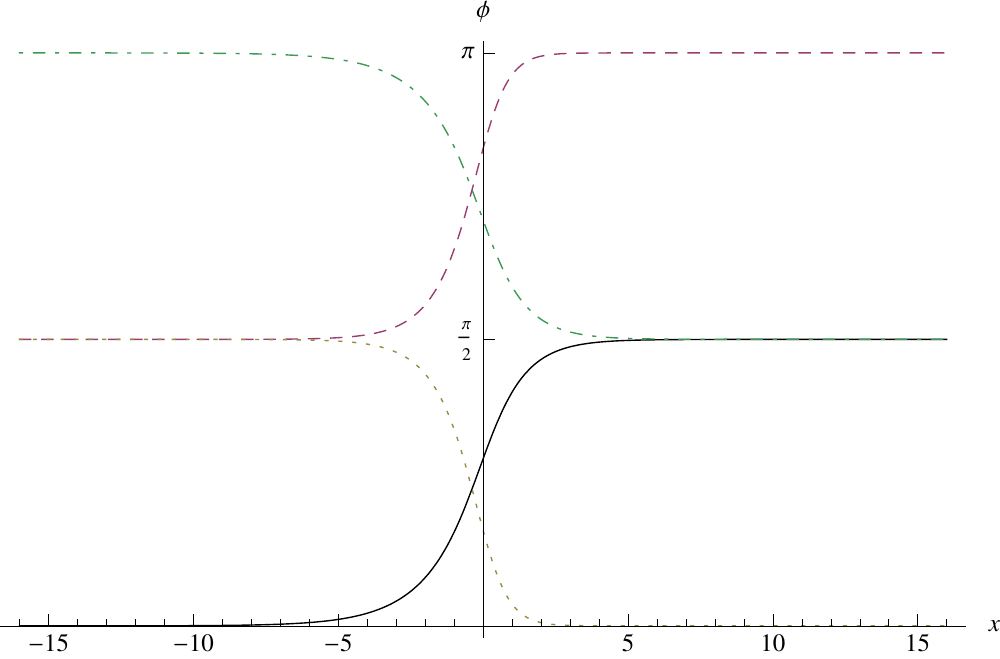}
\caption{There are two types of kinks and anti-kinks. The full and dashed lines are kinks with $l=0,1$ respectively while the  dotted and dot-dashed lines are anti-kinks, again with $l=0,1$ respectively.}
\label{fig:kinks}
\end{figure}

\begin{figure}[!ht]
\centering
\includegraphics[scale=0.5]{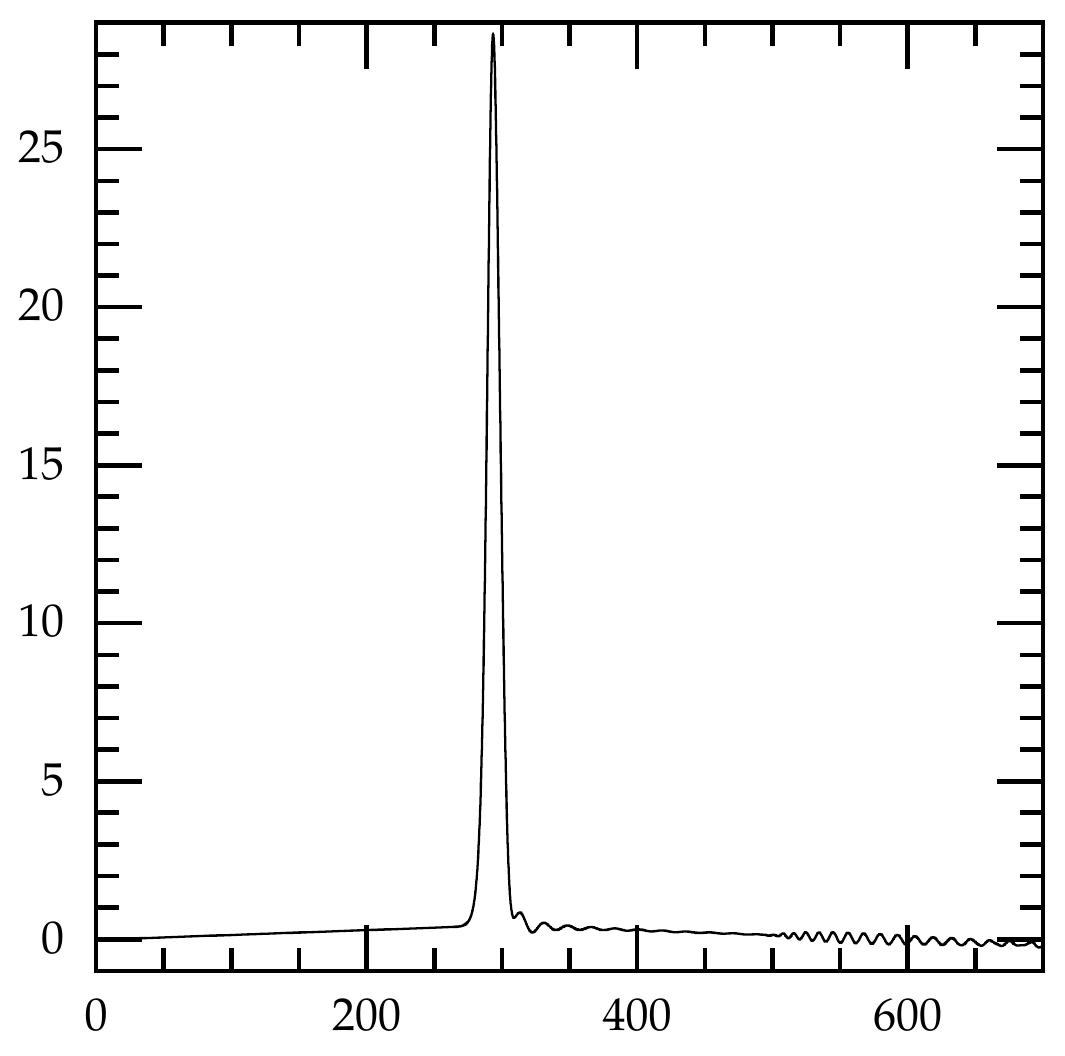}
\caption{Plot of $Q(t)-Q(0)$ against time $t$.}
\label{fig:q_breather}
\end{figure}

The features that we believe are a fingerprint of a quasi-integrable theory were discovered in this model when we considered the scatteing of two kinks and of a kink-antikink. Surprisingly (since the model is definitely non-integrable) the two kinks scattered with very little radiation being emitted and for a kink-antikink (started from rest) 
their configuration lived beyond any expectation behaving like a slowly decaying  breather. This behaviour can be explained by the existence of an infinite number of  charges where the first non-trivial one is given by ($\partial_- \equiv \partial_t-\partial_x$ and $V'\equiv \frac{dV}{d\varphi}$)
$$
Q=\int_{-\infty}^{+\infty}dx\;\left( 4(\partial_-\varphi)^4 + \partial_-\varphi\partial_-^3\varphi + \partial_-^2\varphi \;V' + (\partial_-\varphi)^2 (V'' + 16 V -1) \right)
$$
It is not fully conserved in time, but it is quasi-conserved or asymptotically conserved as showed in figure \ref{fig:q_breather}.

\subsection{The quasi-zero curvature condition}

The behaviour of the observed breather-like structure suggests that the considered model (arising from a deformation of an integrable one) might possess some features that are similar to  the original one. The zero curvature condition provides a way to quantify how much a non-vanishing $\varepsilon$ makes the theory non-integrable. To perform this quantification we can proceed as follows. For a given potential $V$ we first construct the Lax pair potentials as
$$
A_+ = \frac{1}{2}\left( (\omega^2 V - m)b_1 - i\omega \frac{dV}{d\varphi}F_1\right),\qquad A_- = \frac{1}{2}b_{-1}-\frac{i}{2}\omega \partial_- \varphi F_0
$$
with the use of the light-cone coordinates $x_\pm = \frac{1}{2}(t\pm x)$ and the $sl(2)$ loop algebra with generators
$$
b_{2m+1}\equiv \lambda^m(T_+ + \lambda T_-),\quad F_{2m+1}\equiv \lambda^m (T_+-\lambda T_-),\quad F_{2m} \equiv 2\lambda^m T_3.
$$
and
$$
\left[T_3,T_\pm \right]=\pm T_\pm, \qquad \qquad \left[T_+,T_-\right]=2T_3.
$$

Then the compatibility condition for this Lax pair, or the curvature of the connection $A_\mu$, takes the form
\begin{equation}
\label{curv}
F_{+-}=\partial_+A_--\partial_-A_++\left[A_+,A_-\right]=XF_1-\frac{i\omega}{2}\left(\partial^2\varphi + \frac{\partial V}{\partial \varphi}\right)F_0,
\end{equation}
where
\begin{equation}
\label{anomaly}
X=\frac{i\omega}{2}\partial_-\varphi \left(\frac{d^2V}{d\varphi^2}+\omega^2 V - m\right)
\end{equation}
is our anomaly.

Note that the last term in \eqref{curv} vanishes once $\varphi$ is a solution of the equation of motion. In the particular case of $V$ given by \eqref{pot_bazeia}, $X$ vanishes if $\varepsilon=0$, {\it i.e.} for $V=\frac{1}{16}\left(1-\cos(4\varphi)\right)$ ($\omega = 4$, $m=1$). We shall refer to equation \eqref{curv} as the quasi-zero curvature equation.

Using the gauge covariance of \eqref{curv} it is convenient to change to  new Lax potentials $A_\mu \rightarrow a_\mu = gA_\mu g^{-1}-\partial_\mu g g^{-1}$ with $g=\exp\left(\sum_{n=1}^{\infty}\zeta_nF_n\right)$, where $\zeta_n$ are functions to be chosen such that the $F_{n-1}$ components of
\begin{eqnarray*}
A_-&\rightarrow & a_- = \frac{1}{2}b_{-1}-\frac{1}{2}\zeta_1 \left[b_{-1},F_1\right]-\frac{i}{2}\omega \partial_- \varphi F_0 - \frac{1}{2}\zeta_2\left[b_{-1},F_2\right]+\frac{1}{4}\zeta_1^2\left[\left[b_{-1},F_1\right],F_1\right]\\
&-&\frac{i}{2}\omega \partial_- \varphi \zeta_1 \left[F_1,F_0\right]-\partial_-\zeta_1F_1  \dots  -\frac{1}{2}\zeta_n \left[b_{-1},F_n\right]+\dots
\end{eqnarray*}
vanish. As a result the Lax potential  gets rotated into the subalgebra 
\begin{equation}
\label{rot_pot}
a_-=\frac{1}{2}b_{-1}+\sum_{n=0}^{\infty}a_{-}^{(2n+1)}b_{2n+1}
\end{equation}
with
$$
a_-^{(1)}=-\frac{1}{4}\omega^2(\partial_-\varphi)^2, \qquad a_-^{(3)}=-\frac{1}{16}\omega^4(\partial_-\varphi)^4-\frac{1}{4}\omega^2\partial_-^3 \varphi \partial_-\varphi, \dots
$$
with no use of the equations of motion. In a similar way the $A_+$ potential is transformed into $a_+$;  however,
  now equations of motion are used and
and one gets
$$
a_+ = \sum_{n=0}^{\infty}a_+^{(2n+1)}b_{2n+1}+\sum_{n=2}^\infty c_+^{(n)}F_n;
$$
The quantities $c_+^{(n)}$ are proportional to the anomaly $X$ and its derivatives. 

Under this gauge transformation the curvature transforms as $F_{+-}\rightarrow g{F_{+-}}g^{-1}=\partial_+a_- - \partial_-a_++\left[a_+,a_-\right]=XgF_1g^{-1}$ and its components, in the abelian sub-algebra, define the quasi-continuity equation
\begin{equation}
\label{quasi_continuity}
\partial_+a_-^{(2n+1)}-\partial_-a_+^{(2n+1)}=\beta^{(2n+1)}\qquad n=0,1,2,\dots
\end{equation}
with
$$
\beta^{(1)}=0,\qquad \beta^{(3)}=i\omega\partial_-^2\varphi X, \dots
$$
From it, the quantities 
$$
Q^{(2n+1)}\equiv \int_{-\infty}^{+\infty}dx\;a_x^{(2n+1)}, \qquad \qquad \alpha^{(2n+1)}\equiv \int_{-\infty}^{+\infty}dx\;\beta^{(2n+1)},
$$
where $a_x^{(2n+1)}=\frac{1}{2}\left(a_+^{(2n+1)}-a_-^{(2n+1)}\right)$ arise naturally, and we end up with the desired quasi-conservation laws
\begin{equation}
\label{quasi_cons}
\frac{dQ^{(1)}}{dt}=0 \qquad \qquad \frac{dQ^{(2n+1)}}{dt}=-\frac{1}{2}\alpha^{(2n+1)}\qquad n=1,2,\dots
\end{equation}

\subsection{The parity argument}

The asymptotic conservation of the quantities $Q^{(2n+1)}$ defined above takes place when the field configurations satisfy a very specific behaviour under the very special space-time parity transformations $P: (\widetilde{x},\widetilde{t})\rightarrow (-\widetilde{x},-\widetilde{t})$, with $\widetilde{x}=x-x_\Delta$ and $\widetilde{t}=t-t_\Delta$, $x_\Delta$ and $t_\Delta$ being dependent on the parameters of the field configuration. 

Suppose that for such configuration $P(\varphi)=-\varphi +\textrm{const.}$ and $P(V)=V$. This transformation combined with the order two automorphism of the algebra $\Sigma(b_{2n+1})=-b_{2n+1}$, $\Sigma(F_{2n})=-F_{2n}$, $\Sigma(F_{2n+1})=F_{2n+1}$ leads to the mirror-like property of the charges $Q^{(2n+1)}$:
\begin{equation}
\label{mirror}
Q^{(2n+1)}(t=\widetilde{t}_0+t_\Delta)=Q^{(2n+1)}(t=-\widetilde{t}_0+t_\Delta).
\end{equation}

Thus, what determines the asymptotic conservation or the quasi-conservation of the charges are the properties of the anomaly $\alpha^{(2n+1)}$. In order to evaluate this quantity we  perform an expansion of the equation of motion, as well as of the field configuration (and all parameters), in powers of $\varepsilon$:
$$
\varphi = \varphi_0 + \varepsilon \varphi_1 + \varepsilon^2 \varphi_2 + \cdots \qquad
V=V\vert_{\varepsilon = 0}+\frac{dV}{d\varepsilon}\vert_{\varepsilon=0}\varepsilon+\cdots
$$
At order zero one gets the integrable field equation $\partial^2 \varphi_0 + \frac{1}{4}\sin(4\varphi_0)=0$. At first order, $\varphi_1$ has to satisfy the equation of motion
$$
\partial^2 \varphi_1 + \cos(4\varphi_0)\varphi_1 = \sin(\varphi_0)\cos(\varphi_0)\left[2\sin^2(\varphi_0) \ln(\sin^2(\varphi_0))+\cos^2(\varphi_0)\right].
$$

Next we have to analyse the behaviour of all these expressions under the parity transformation. We know that the sine-Gordon exact two-soliton solution $\varphi_0$ is odd under $P$, {\it i.e.}, $P(\varphi_0)=-\varphi_0 + \pi$. Then, for the next order terms, we introduce $\varphi_1^{(\pm)}=\frac{1}{2}\left(1\pm P\right)\varphi_1$, corresponding to the even and odd parts of the field configuration under our parity transformation. Then we find that the odd parts $\varphi^{(-)}_1$ satisfy linear non-homogeneous equations and that the even parts $\varphi^{(+)}_1$ satisfy linear homogeneous equations. This suggests that for the even part there is the trivial solution $\varphi^{(+)}_1=0$, while this is not the case for the odd part. Then, if $\varphi_1$ is a solution, so is $\varphi_1 - \varphi_1^{(+)}$, which is odd under $P$. This argument can be extended to all orders of perturbation in $\varepsilon$.   The anomaly for these odd two-soliton solutions 
vanishes in all orders of perturbation and therefore we have our quasi-conservation property.

\section{Numerical support}

The quasi-conservation law discussed above  has also been seen in numerical simulations in the scattering of two kinks, kink and anti-kink and in a system with two kinks and an anti-kink. To perform their time evolution the fourth-order Runge-Kutta method was used with a lattice of 10001 equally spaced points with size of around 30 times the size of the kink. Specific boundary conditions were used in order to absorb the radiation at the borders (not to have any reflections).

\begin{figure}[!ht]
\centering
\includegraphics[scale=0.5]{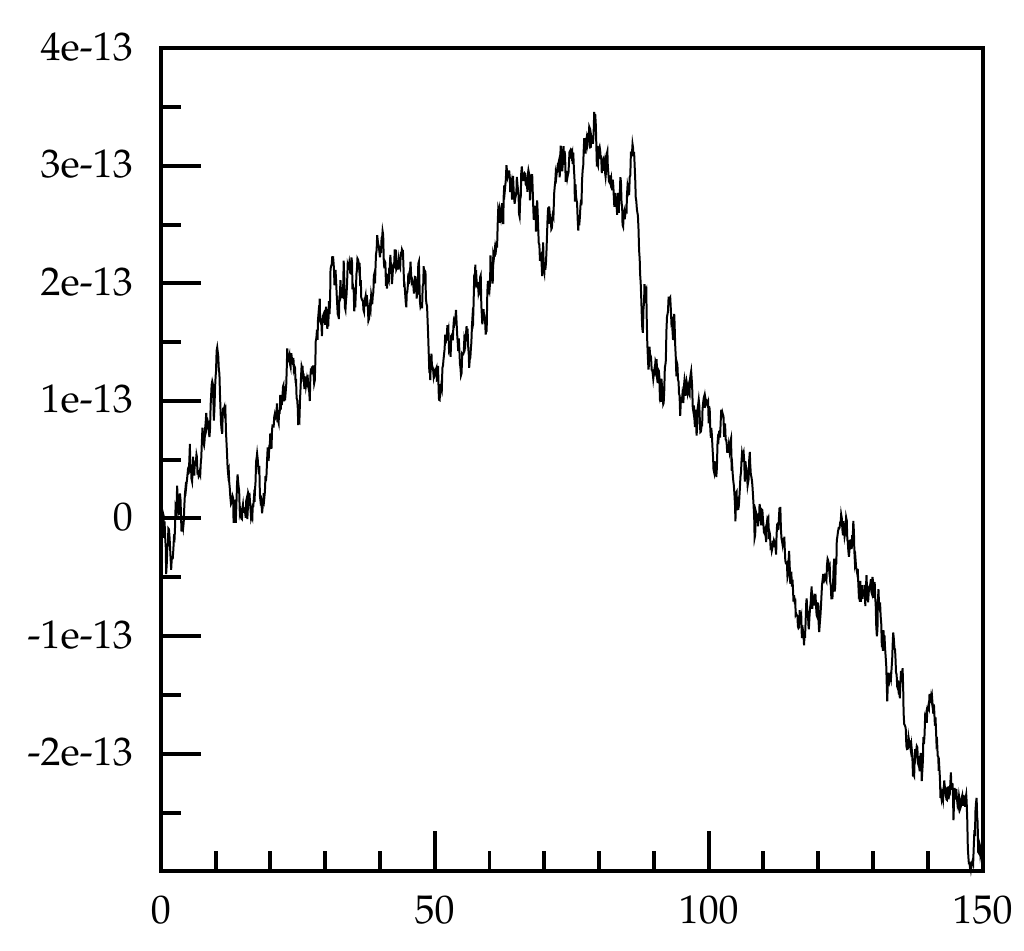}
\caption{$Q^{(3)}(t)-Q^{(0)}(t)$ for $\varepsilon = 0.0$}
\label{fig:num_1}
\end{figure}

Below we present the values obtained for $Q^{(3)}(t)-Q^{(0)}(t)$ from these simulations. Figure \ref{fig:num_1} corroborates the fact that the anomaly for the integrable sine-Gordon model ($\varepsilon = 0$) vanishes.

\begin{figure}[!ht]
\centering
\includegraphics[scale=0.5]{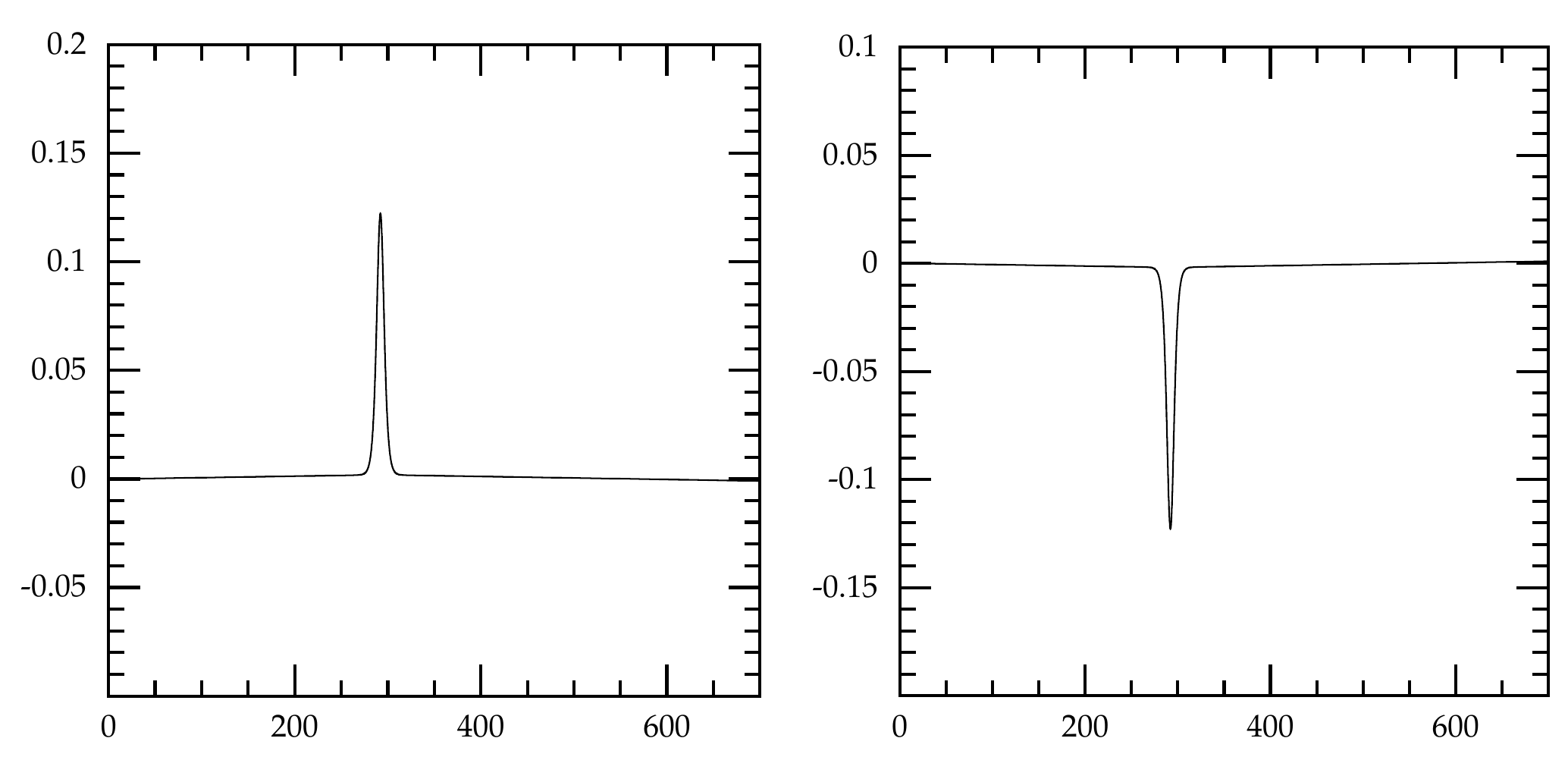}
\caption{$Q^{(3)}(t)-Q^{(0)}(t)$ for $\varepsilon = -0.01$ and $\varepsilon=0.01$.}
\label{fig:num_2}
\end{figure}

\begin{figure}[!ht]
\centering
\includegraphics[scale=0.5]{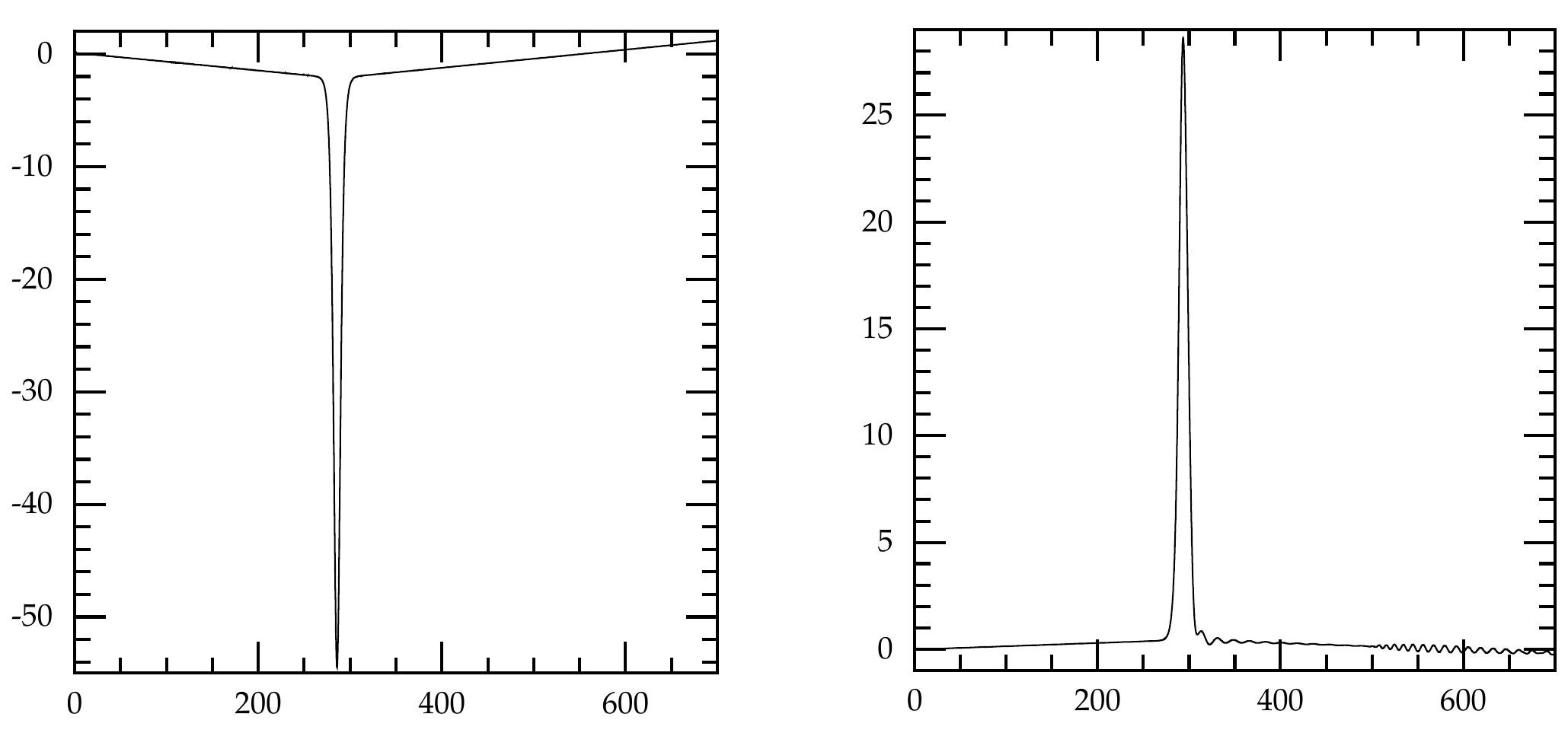}
\caption{$Q^{(3)}(t)-Q^{(0)}(t)$ for $\varepsilon = -1.9$ and $\varepsilon=3.5$.}
\label{fig:num_3}
\end{figure}

For other values of $\varepsilon \neq 0$ we see the asymptotic conservation, as predicted analytically. In figure \ref{fig:num_2} we present
the plots of the values of $Q^{(3)}(t)-Q^{(0)}(t)$ for $\varepsilon$ very close to zero, and in figure \ref{fig:num_3} for values of $\varepsilon$ not so close.

%

\section{Final remarks}

The deformation of the sine-Gordon model considered here \cite{SG} allows certain field configurations to behave in a fashion similar to those of the integrable theory. This may happen for very many deformations and we hope that quasi-integrability may be a new non-linear phenomenon. Once the methods  to define it properly and to understand it are setted, and we have started doing this through the study of this model (and on the way adapting the well-known algebraic techniques from the study of solitons) quasi-integrability may become as useful as integrability itself, leading to a wide range of applications, specially because of the possibility of dealing with cases which lie outside the areas covered by integrability. The main ingredient in our understanding of  quasi-integrability at the moment is the parity transformation; however, it is not clear yet whether this property is fundamental or descendent from more general principles. Or for that matter what really are  the physical principles behind the tuning of the properties of the field configuration giving the desired behaviour under parity. Recently, we have also considered  deformations of the integrable non-linear Schr\"odinger model \cite{NLS} and the method developed here was there also successfully applied. This suggests to us that we are on a promising track. But the real next step will involve generalising our  approach to study quasi-integrability in higher ({\it i.e.} more physical)  dimensions.

\begin{theacknowledgments}

LAF is grateful to the organizes of {\em 2nd International Workshop on Nonlinear and Modern Mathematical Physics} for the invitation to give this talk and for the financial support. LAF is partially supported by CNPq-Brazil. 
The present manuscript was completed when  LAF and WJZ  visited the Mathematisches Forschungsintitut in Oberwolfach (MFO). They would like to thank the MFO for its hospitality.
  
\end{theacknowledgments}



\bibliographystyle{aipproc}   

\bibliography{sample}

\IfFileExists{\jobname.bbl}{}
 {\typeout{}
  \typeout{******************************************}
  \typeout{** Please run "bibtex \jobname" to optain}
  \typeout{** the bibliography and then re-run LaTeX}
  \typeout{** twice to fix the references!}
  \typeout{******************************************}
  \typeout{}
 }


\end{document}